\documentclass[12pt]{article}

\usepackage{aasms4}

\def\etal{{\it et al.}}
\def\ie{{\it i.e.}}
\def\eg{{\it e.g.}}
\def\lap{\hbox{${_{\displaystyle<}\atop^{\displaystyle\sim}}$}}
\def\gap{\hbox{${_{\displaystyle>}\atop^{\displaystyle\sim}}$}}

\begin{document}

\title{Starquake-induced Magnetic Field and Torque Evolution in
Neutron Stars}

\author{Bennett Link\altaffilmark{1}}
\affil{Montana State University, Department of Physics, Bozeman MT
59717; blink@dante.physics.montana.edu} 
\altaffiltext{1}{Also Los Alamos National Laboratory}
\author{Lucia M. Franco}
\affil{University of Chicago, 5640 S. Ellis Ave., Chicago IL 60637; 
lucia@oddjob.uchicago.edu}
\and
\author{Richard I. Epstein}
\affil{Los Alamos National Laboratory, Mail Stop D436, Los Alamos, NM
87545; epstein@lanl.gov}

\begin{abstract}

The persistent increases in spin-down rate ({\em offsets}) seen to
accompany glitches in the Crab and other pulsars 
suggest increases in the spin-down torque.  We interpret these offsets
as due to {\em starquakes} occurring as the star spins down and the
rigid crust becomes less oblate. We study the evolution of strain in
the crust, the initiation of starquakes, and possible consequences for
magnetic field and torque evolution. Crust cracking occurs as
equatorial material shears under the compressive forces arising from
the star's decreasing circumference, and matter moves to higher
latitudes along a fault inclined to the equator.  A starquake is most
likely to originate near one of the two points on the rotational
equator farthest from the magnetic poles. The material breaks along a
fault approximately aligned with the magnetic poles.  We suggest that
the observed offsets come about when a starquake perturbs the star's
mass distribution, producing a misalignment of the angular momentum
and spin axes.  Subsequently, damped precession to a new rotational
state increases the angle $\alpha$ between the rotation and magnetic
axes. The resulting increase in external torque appears as a permanent
increase in the spin-down rate. Repeated starquakes would continue to
increase $\alpha$, making the pulsar more of an orthogonal rotator.

\end{abstract}

\keywords{dense matter --- magnetic fields --- stars: magnetic fields ---
stars: neutron --- pulsars: individual (Crab)}

\section{Introduction}

The magnetic braking torque acting on an isolated neutron star would
be steady in the absence of abrupt changes to the star's magnetic
configuration. Most pulsars, however, do not slow in a regular
fashion, but suffer variations in their spin rates in the form of
glitches and timing noise. Perhaps the most striking aspect of spin
evolution is the persistent increases in spin-down rate that accompany
glitches in the Crab pulsar (\cite{LPS}), PSR 0355+54 (\cite{lyne87})
and PSR 1830-08 (\cite{sl96}); see Table 1.  In the Crab, these
permanent {\em offsets} involve fractional changes in the spin down
rate of $\sim 10^{-6}-10^{-4}$ (see Fig. 1). PSR 1830-08 has exhibited
one offset of $8\times 10^{-4}$, and a persistent offset might have
followed the large 1986 glitch of PSR 0355+54. Similar offsets
might also occur in the Vela pulsar (\cite{le97}), giving rise to the
small braking index of $1.4\pm 0.2$ reported by Lyne \etal\ (1996).

It is striking that all observed offsets are of the same sign, and
correspond to {\em increases} in the spin-down rate.  One
interpretation of this phenomenon is that glitches are accompanied by
sudden and permanent increases in the external torque acting upon the
star (\cite{leb92}; \cite{le97}). Such torque changes could occur if
either the direction or magnitude of the star's magnetic moment
changes. {\em Starquakes}, occurring as the star spins down
(\cite{starquakes}; \cite{ruderman76}) would affect the external
torque if they change the orientation of the magnetic moment with
respect to the spin axis. If structural relaxation occurs
asymmetrically about the rotation axis, due perhaps to magnetic
stresses or asymmetric material properties, the star's spin and
angular momentum vectors would become misaligned. As the star
precesses and relaxes to a new rotational state, the magnetic moment
would assume a new orientation with respect to the rotation axis,
leading to a change in the external torque. In this paper, we study
how the rigid neutron star crust relaxes its structure, and consider
possible consequences for evolution of the magnetic field and torque. 

Alpar and Pines (1993; also \cite{more_capacitors}) have suggested
that the Crab's offsets result from a reduction in the moment of
inertia on which the external torque acts. Such a change could occur
by either a structural change of the star, \eg, the star becomes less
oblate, or through a decoupling of a portion of the star's liquid
interior from the external torque.  If the moment of inertia decreases
through structural readjustment, to conserve angular momentum, the
star would always spin {\em more} rapidly than had the glitch not
occurred (\cite{leb92}; \cite{le97}). The large offsets following the
1975 and 1989 glitches in the Crab, however, eventually caused the
star to spin {\em less} rapidly than had the glitch not occurred
(\cite{LPS}); the observed offsets, therefore, cannot be due {\em
solely} to structural readjustments. In principle, the Crab's
spin-down rate offsets could be due to decoupling of a portion of the
star's superfluid interior from the external torque
(\cite{capacitors}; \cite{more_capacitors}), though quantitative
agreement of this model with the data has yet to be demonstrated.
Torque increases associated with the surface field structure appears
to be the most straightforward explanation.

A neutron star relaxes its oblateness as it spins down, moving
equatorial material toward the rotation axis and reducing the
equatorial circumference. If the stellar crust is brittle, stresses
lead to {\em starquakes} as the yield strength of the crustal material
is exceeded (\cite{starquakes}).\footnote{Superfluid stresses in the
crust (\cite{ruderman76}) could also drive crust cracking, but will
not be considered here.} The actual response of solid neutron-star
matter at high pressure to shear is not well-understood. Here we
assume that the neutron star crust is brittle and explore the
consequences of this assumption.\footnote{ Known materials exhibit
ductile rather than brittle behavior under pressures comparable to
their material shear moduli (\cite{duba90}). Nevertheless, deep-focus
earthquakes are known to originate from regions of very high pressure
(\cite{gh95}). These faults are thought to be facilitated by
densification phase changes; small regions of higher density nucleate
as the material is stressed, and act as a lubricant for shearing
motion. Analogous processes might occur in the high-pressure material
of the neutron star.}

\section{Growth of Strain in a Spinning-down Star}

To study the qualitative features of the growth of strain in a neutron
star as it spins-down under its external torque, we model the star as
an incompressible, homogeneous, self-gravitating elastic sphere of
constant shear modulus.  The non-rotating configuration is a sphere of
radius $R$. We take the initial configuration to be unstrained with
rotation frequency $\Omega$, and investigate how the star deforms as
the spin rate changes by $-\delta\Omega$ where $0<\delta
\Omega\ll\Omega$. The initial rotation produces an equatorial bulge of
relative height $\sim (R\Omega/v_k)^2\ll 1$, where $v_k =
(2GM/R)^{1/2}$ is the Keplerian velocity near the stellar surface. A
change $-\delta\Omega$ in the rotation rate produces a small decrease
in the equatorial bulge of $\sim\Omega\delta\Omega R^2/v_k^2$. As the
star deforms, matter originally at $\vec r$ is displaced to $\vec r +
\vec u(\vec r)$, where $\vec u(\vec r)$ is the {\em displacement
field}. The displacement field can be found using elasticity theory
(see, \eg, \cite{love20}; \cite{ll}). In a spherical coordinate system
centered on the star, with $\theta=0$ at the star's north pole, the
displacement field is (see, also, \cite{starquakes})
\begin{eqnarray}
\label{u}
u_r (r,\theta) & = &  {\Omega\delta\Omega R^2\over 
v_k^2} r\left (r^2 -{8\over 3}\right ) \left (1 - 3 \cos^2 \theta
\right ) \cr
u_\theta (r,\theta) & = & {\Omega\delta\Omega R^2\over 
v_k^2} r\left (5r^2 - 8\right ) \sin\theta\cos\theta. 
\end{eqnarray}
Here $r$ and $u$ are in units of the stellar radius, and we have
neglected terms of order $(c_t/v_k)^2\lap 10^{-4}$ where $c_t$ is
the transverse sound speed. The displacement field is illustrated in
Fig. 2.

The local distortion of the solid is described by the
strain tensor
\begin{equation}
\label{strain_tensor}
u_{ij} = {1\over 2}\left ({\partial u_i\over\partial x_j} + 
                          {\partial u_j\over\partial x_i}\right ). 
\end{equation}
In a local coordinate system in which this matrix is diagonal, the
eigenvalues $\epsilon_i$ represent compression ($\epsilon_{i}<0$)
or dilation ($\epsilon_{i}>0$) along the respective axes.  Suppose the
largest and smallest eigenvalues are $\epsilon_l$ and $\epsilon_s$,
respectively. The {\em strain angle} is $\epsilon_l -
\epsilon_s$, and the plane containing the corresponding principal axes is the
{\it stress plane}. For the uniform sphere, we find that the stress
plane is tangential to the stellar surface near the equator with
compression in the azimuthal direction and dilation toward the
rotational poles (see Fig. 3). Near the poles, the stress plane
switches to the radial-azimuthal plane, \ie, along cones.  These
results are consistent with those of Baym and Pines (1971), who
calculated the strain field as a sphere at zero pressure spins down.

\section{Starquakes}

At zero pressure, a stressed element of matter can break due to shear
or tension. At the pressure of the neutron star crust ($p\gg\rho
c_t^2$), however, only the former process can occur. Compressed
material strained at a critical angle shears along a plane as shown in
Fig. 4 (\cite{gh95}). The angle this plane takes with respect to the
plane of maximum compression depends on the material's internal
friction, and is in the range $30-45^\circ$. The critical strain angle
of neutron star material is not accurately known. Highly-compressed
terrestrial minerals can support strain angles of $\gap 0.01$
before breaking (\cite{billings72}; \cite{duba90}).

The geometry of a starquake is independent of the value of the
critical strain angle for neutron star matter.  Strain builds as the
star slows and reduces its circumference.  For the homogeneous sphere,
the maximum strain is reached on the equator, as each material element
is compressed by its neighbors on the equator (see Fig. 3). When the
strain angle for an element reaches a critical value, the material
breaks along a fault that takes an angle of $30-45^\circ$ with respect
to the equator and propagates to higher latitudes (see Fig. 5).
Movement of material from the equatorial bulge to higher latitudes
decreases the star's oblateness.  At higher latitudes the strain angle
diminishes and the fault terminates (\cite{hertzberg96}).  This
picture of starquakes differs fundamentally from that of Baym \&\
Pines (1971) who considered the distortion in a sphere at zero
pressure.  They found that the material breaks as tensile forces pull
apart material in the equatorial plane.  In our analysis, however, the
enormous pressure in neutron star matter prevents tensile rupturing.

For an axially-symmetric star, the crust is equally likely to begin
breaking at every point on the equator. The rotation-induced
displacements of the star's crust distort the initial magnetic field
$\vec B (\vec r)$ that is anchored in the stellar crust thus
generating magnetic stresses that affect the development of strain. 
The strength of the magnetic stress relative to the shear stress is
$\sim\beta \equiv (v_A/c_t)^2 $, where $v_A$ is the Alfv\'en
velocity. Through most of the crust the induced magnetic stresses are
significantly smaller than the material shear stresses and can be
treated perturbatively. For example, at moderate densities in the
outer crust we have $\beta =8 \times 10^{-4}(B/10^{12} {\rm G})^2
(\rho/10^{10}{\rm\, g\, cm}^{-3})^{-1}(c_t/ 10^8 {\rm cm}\ {\rm
s}^{-1})^{-2}$, while the magnetic stresses can exceed the material
ones at low densities, below $\sim 10^7{\rm\, g\, cm}^{-3}$.

The strain field that develops in magnetized material could be
obtained by solving the elasticity equations and Maxwell's equations
self-consistently, subject to boundary conditions at the star's
surface. To illustrate how the star's magnetic field determines where
starquakes originate, we consider the simpler problem of how the
presence of a magnetic field affects the material strain in an
infinite medium. In the absence of the field, the equilibrium state is
given by
\begin{equation}
{\partial\sigma_{ij}\over\partial x_i} + F_j = 0,
\end{equation}
where $F_j$ represents gravitational and centrifugal forces. In the
presence of a magnetic field, the stress tensor changes by $\delta\sigma_{ij}$.
The displacement of the material induces a change in the field
$\delta\vec B$, with a corresponding change in the Maxwell stress
tensor $T_{ij}$ of $\delta T_{ij}$. We take the unperturbed field
$\vec B$ to be dipolar, which gives no force on the undistorted
matter, \ie, $\partial T_{ij}/\partial x_i=0$. The new equilibrium is
given by
\begin{equation}
{\partial\over\partial x_i}(\delta\sigma_{ij} + \delta T_{ij}) = 0.
\end{equation}
For an infinite medium (one in which boundary effects are
unimportant), the solution is $\delta\sigma_{ij}=-\delta T_{ij}$. The
magnetic field induces a change $\delta\vec u$ in the displacement
field. For a field that is sufficiently small that this correction to
the displacement field is small, the perturbation to the Maxwell
stress tensor is
\begin{equation}
\delta T_{ij} = {1\over 4\pi}\left [\delta B_i B_j + 
        B_i\delta B_j - B_{k}\delta B_{k}\delta_{ij}\right ],
\end{equation}
with 
\begin{equation}
\delta\vec B = \nabla\times (\vec u\times \vec B). 
\end{equation}
Here $\vec u$ is the displacement field in the absence of the field. 
For incompressible matter, the correction $\delta u_{ij}$ to the
strain tensor is
\begin{equation}
\delta u_{ij} = {1\over 2\mu}\left (\delta\sigma_{ij} - {1\over 3}
              \delta\sigma_{ll}\right ), 
\end{equation}
where $\mu$ is the shear modulus. The eigenvalues of this tensor give
the correction to the strain angle due to the induced magnetic forces.
Using the displacement field of eq. [1] in eq. [6], we obtain an
estimate of the corrections to the strain angle on the star's equator;
the result is shown in Fig. 6. Near the magnetic poles, the field
gives the material extra rigidity, and reduces the strain
angle. Consequently, as the star spins down, the strain increases the
fastest at the two equatorial points farthest from the poles.  These
results indicate that a starquake will begin at one of these two
points when critical strain is reached. A more complete treatment
would account for magnetic stresses at the stellar surface; we do not
expect treatment of the boundary to alter our qualitative conclusions.

Isotropic material under compressive stress is as likely to break
along the plane indicated in Fig. 4 as along its complementary plane.
In a neutron star, however, the magnetic field gives the material
extra rigidity along shear planes that cross field lines. Cracking is
thus favored along the plane that is most parallel to the field (fault
$f$ in Fig. 5). 

\section{Effects on the Spin-down Torque}

In the picture of starquakes described above, a slowing star reduces
its equatorial circumference by shearing material across the equator,
moving the material along the fault to higher latitudes (see Fig. 5).
We now discuss how this crustal motion could affect the spin-down
torque acting on the star by producing a misalignment of the star's
spin and angular momentum vectors. The star then undergoes damped
precession to a new rotational state.\footnote{Crust motion would also
change the external torque by moving the magnetic poles with respect
to the rotation axis. The associated change in $\alpha$, however, is
only $\sim\Delta R/R\sim 10^{-5}$, too small to account for the
largest persistent shifts seen in the Crab.}

Quake-induced mass motion produces a small change in the orientation
of the principal axes of the star's moment of inertia.  The material
motions create ``mountains'' ($\sim 100$ $\mu$m high) at higher
latitudes, breaking the axial symmetry of the star's mass
distribution. Suppose that before the starquake the largest principal
moment of inertia corresponds to a principal axis aligned with the
rotation axis. As the mass motion occurs, the principal axis of
inertia shifts {\em away} from the magnetic axis by $\Delta\alpha$,
pointing in a new direction $e_1$ (fixed in the star, see Fig. 5).  The star's
rotation and angular momentum vectors are now misaligned and the
rotation axis precesses about the angular momentum axis
(\cite{shaham77}).  Dissipative processes, such as the interaction of
superfluid vortices with the nuclei of the inner crust, damp the
precessional motion (\cite{swc}), bringing the star to a new stable
equilibrium in which the principal axis of inertia is again aligned
with the star's angular momentum.  As the precessions damps, the angle
between the star's rotation and magnetic axes $\alpha$ {\em increases}
by $\Delta\alpha$.  In some models of pulsar spin down, an increase in
$\alpha$ leads to an increase in the spin-down torque.  In the
magnetic dipole model for pulsar slowdown, for example, an increase by
$\Delta\alpha$ gives a {\em permanent} increase in the spin-down rate
of $\Delta\dot\Omega/\dot\Omega=2\Delta\alpha/\tan\alpha$. We suggest
that the Crab's offsets arise from starquake-induced changes of order
$\Delta\alpha\sim 10^{-4}$ (\cite{le97}).

The following dimensional argument gives an estimate of the magnitude
of the shift of the principal axes.  Suppose that between starquakes
the star spins down by $\delta\Omega$ and reduces its equatorial
circumference by $\Delta R$. As the starquake occurs, crust material
suddenly moves along the fault by a comparable amount. From eq.
[\ref{u}], we estimate
\begin{equation}
{\Delta R\over R} 
\simeq {10\pi R^2 \Omega\delta\Omega\over 3v_k^2} =
4\times 10^{-5} \left ({\Omega\over 200\ {\rm rad}\ {\rm s}^{-1}}\right )^2 
\left ({v_k\over 10^{10}\ {\rm cm}\ {\rm s}^{-1}}\right )^{-2} 
\left ({t_q\over 14\ {\rm yr}}\right )
\left ({t_{\rm age}\over 10^3\ {\rm yr}}\right )^{-1}, 
\end{equation}
where $t_q$ is the time interval between large quakes and $t_{\rm age}$ is
the spin-down age.  Assuming that quakes are coincident with glitches,
$t_q\sim 14$ yr for the Crab pulsar. 
Matter accumulates in mountains over a length
scale of lateral extent comparable to the fault length $L$. We expect
$L$ to be roughly the length scale characterizing the strain field, of
order $R$.  Upon formation of mountains,
$\hat e_1$ changes its angle with respect to the rotation axis by an
amount of order the ratio of the off-diagonal elements of the moment
of inertia tensor to the bulge moment of inertia; the shift is of
order $\Delta\alpha\sim (M_m/M_B)(L/R)$, where $M_m$ is the mass of a
mountain and $M_B$ is the mass in the equatorial bulge.  The mass in
the bulge is of order $M_B\sim (R\Omega/v_k)^2 M_c$, where $M_c$ is
the mass of the crust.  The fraction of crust mass that accumulates in
a mountain is roughly the fraction of the star's area that moves with
the fault, \ie, $M_m\sim(\Delta R L/R^2)M_c$. Hence, we obtain an
upper limit of 
\begin{equation}
\label{delta-a}
\Delta\alpha \lap {10\pi\over 3}{L^2 \delta\Omega\over R^2\Omega} = 
0.07\left ({L\over R}\right )^2 
\left ({t_q\over 14\ {\rm yr}}\right ) \left ({t_{\rm age}
\over 10^3\ {\rm yr}}\right )^{-1}. 
\end{equation} 
For the Crab pulsar, Rankin (1990) obtains $\alpha=86^\circ$ from
radio data, and Yadigaroglu \& Romani (1995) find $\alpha=80^\circ$
from their analysis of gamma-ray and radio data.  The corresponding
fractional change in spin-down rate (=$2\Delta\alpha/\tan\alpha$) is
$\lap 10^{-2}$ for $\alpha=80^\circ$, large enough to explain the
observed persistent shifts; $\alpha=86^\circ$ gives a limit of
$2\times 10^{-2}$. We cannot yet apply eq. [\ref{delta-a}] to PSRs
0355+54 and 1830-08; only one large glitch has been observed in each
of these pulsars, so the quake intervals and typical offset magnitudes
are unknown. Moreover, the magnitude of the permanent offset in PSR
0355+54, if any, is unknown since the spin-down rate has continued to
changed following the large glitch of 1986.

\section{Summary}

The persistent increases in spin-down rate ({\em offsets}) seen to
accompany glitches in the Crab and other pulsars 
suggest sudden increases in the spin-down torque.  Starquakes
occurring as a neutron star spins-down and readjusts its structure
affect the spin-down torque exerted on the star by changing the
magnetic field geometry or orientation. In this paper we have examined
the evolution of strain in the crust of a spinning-down neutron star
and the initiation of starquakes as the material reaches critical
strain.  Crust cracking occurs as equatorial material shears under the
compressive forces arising from the star's decreasing circumference.
The star decreases its oblateness as matter is moved to higher
latitudes along a fault inclined to the equator.  Magnetic stresses
suppress shearing near the magnetic poles, especially shearing motions
across the field lines.  Starquakes are thus most likely to originate
near the two points on the equator farthest from the magnetic poles
and propagate toward the magnetic poles.

Starquake-induced misalignment of the star's angular momentum and
spin, associated with glitches, is a possible explanation for the
spin-down offsets seen in the Crab pulsar. Following the misalignment,
damped precession to a state of larger angle $\alpha$ between the
magnetic and rotation axes could increase the external torque, giving
a permanent increase in the spin-down rate. Repeated starquakes would
continue to increase $\alpha$, making the pulsar more of an orthogonal
rotator.

\acknowledgements

We thank A. Olinto, P. Richards and D. Longcope for valuable
discussions. This work was performed under the auspices of the U.S.
Department of Energy, and was supported in part by NASA EPSCoR Grant
\#291748, and by IGPP at LANL.

\newpage

\hspace*{4.5cm} TABLE 1 \\[.5cm]
\begin{tabular}{lllll} \hline\hline
PSR & age (yr) & glitch year & $\Delta\Omega/\Omega$ & permanent offset $\Delta\dot\Omega/|\dot\Omega|$ \\ \hline\hline
Crab$^1$ & $10^3$ &\ \ 1969 & $4\times 10^{-9}$ & \ \ $-4\times 10^{-6}$ \\
     & &\ \ 1975 & $4\times 10^{-8}$ & \ \ $-2\times 10^{-4}$ \\
     & &\ \ 1986 & $4\times 10^{-9}$ & \ \ $-2\times 10^{-5}$ \\
     & &\ \ 1989 & $9\times 10^{-8}$ & \ \ $-4\times 10^{-4}$ \\
0355+54$^2$& $6\times 10^5$ &\ \ 1986$^3$  & $4\times 10^{-6}$ & \ \ 
$-3\times 10^{-3}${\bf?}\\
1830-08$^2$& $2\times 10^5$ &\ \ 1990 & $2\times 10^{-6}$ & \ \ $-8\times 10^{-4}$ \\ \hline
\end{tabular}
 
\noindent 
$^1$Lyne, Pritchard \& Smith 1993; $^2$Shemar \& Lyne 1996\\
\noindent
$^3$The magnitude of this offset is quite uncertain, as the pulsar has
not had time to completely recover from the glitch.

\newpage

\figcaption{Twenty-five years of spin history of the Crab pulsar
showing apparent torque increases (adapted from Lyne, Pritchard, \&\
Smith 1993).  Shown are spin rate residuals relative to a model for
data prior to the first glitch.  Glitches are indicated by arrows.
Following each glitch, the pulsar acquired a greater spin-down rate
than before the glitch. The small glitch of 1986 was also followed by
a small offset, shown by the dashed line. These offsets appear to be
permanent and cumulative.}

\figcaption{The displacement field in a spinning-down neutron star.
A cross-section through the center of the star is shown.  The
equatorial diameter decreases, while the polar diameter increases. }

\figcaption{Strain eigenvalues and strain angle vs. latitude
$\theta$ at the stellar surface for $B=0$ (arbitrary units). The
short-dashed curves give the strain angle; the other curves are the
eigenvalues.}

\figcaption{Breaking of a compressed block of matter. A block under
horizontal compression and vertical tension shears along a plane as
shown when the critical strain is reached. Shearing along a
complementary plane flipped over with respect to the plane shown is
equally likely for isotropic material.  The $y-z$ plane is the {\em
stress plane}.}

\figcaption{A starquake. In the absence of a magnetic field, the
material is equally likely to begin breaking along faults $f$ and
$f^\prime$, anywhere on the equator. In the presence of magnetic
stresses, fault $f$ is more likely, creating ``mountains'' (indicated
by the snow-capped peaks) and shifting the largest principal axis of
inertia to a new direction $e_1$ (fixed in the star).}

\figcaption{The strain angle of surface equatorial material in the
presence of a magnetic field (arbitrary units). The curves correspond
to different values of the angle $\alpha$ between the magnetic and
rotation axes. The magnetic poles are at azimuthal angles $\pi/2$ and
$3\pi/2$.}

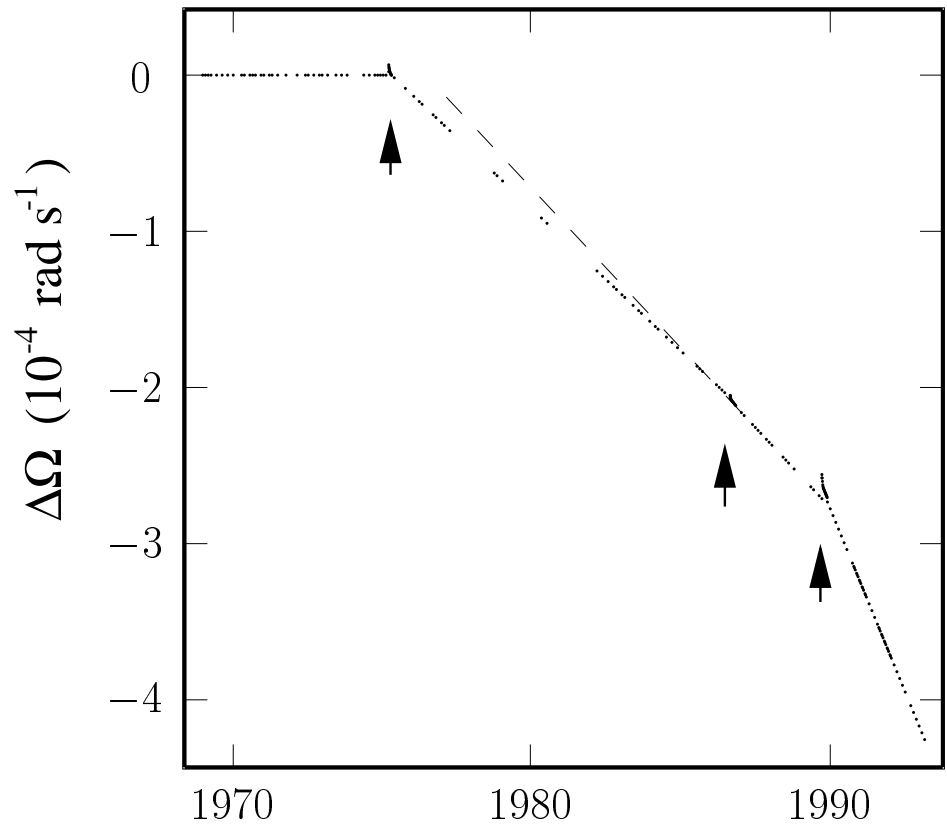
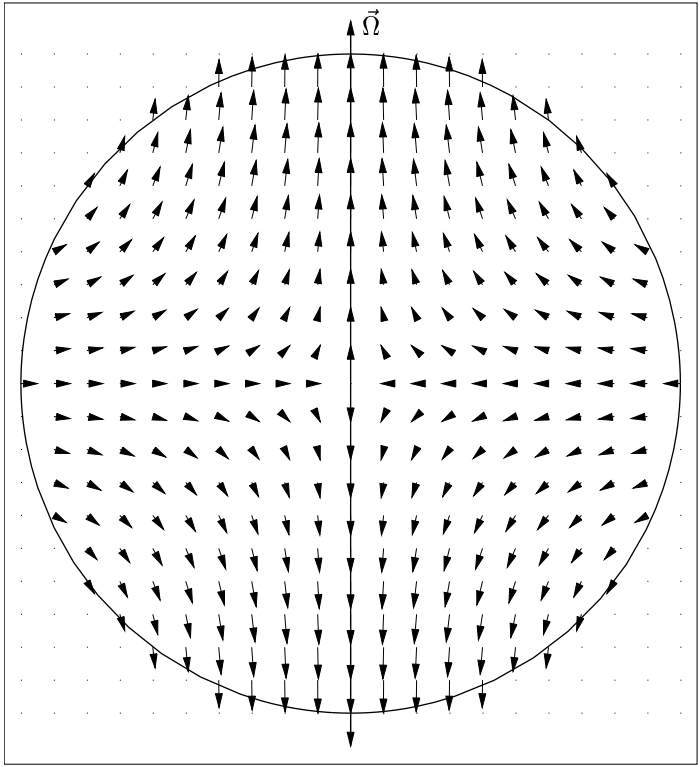
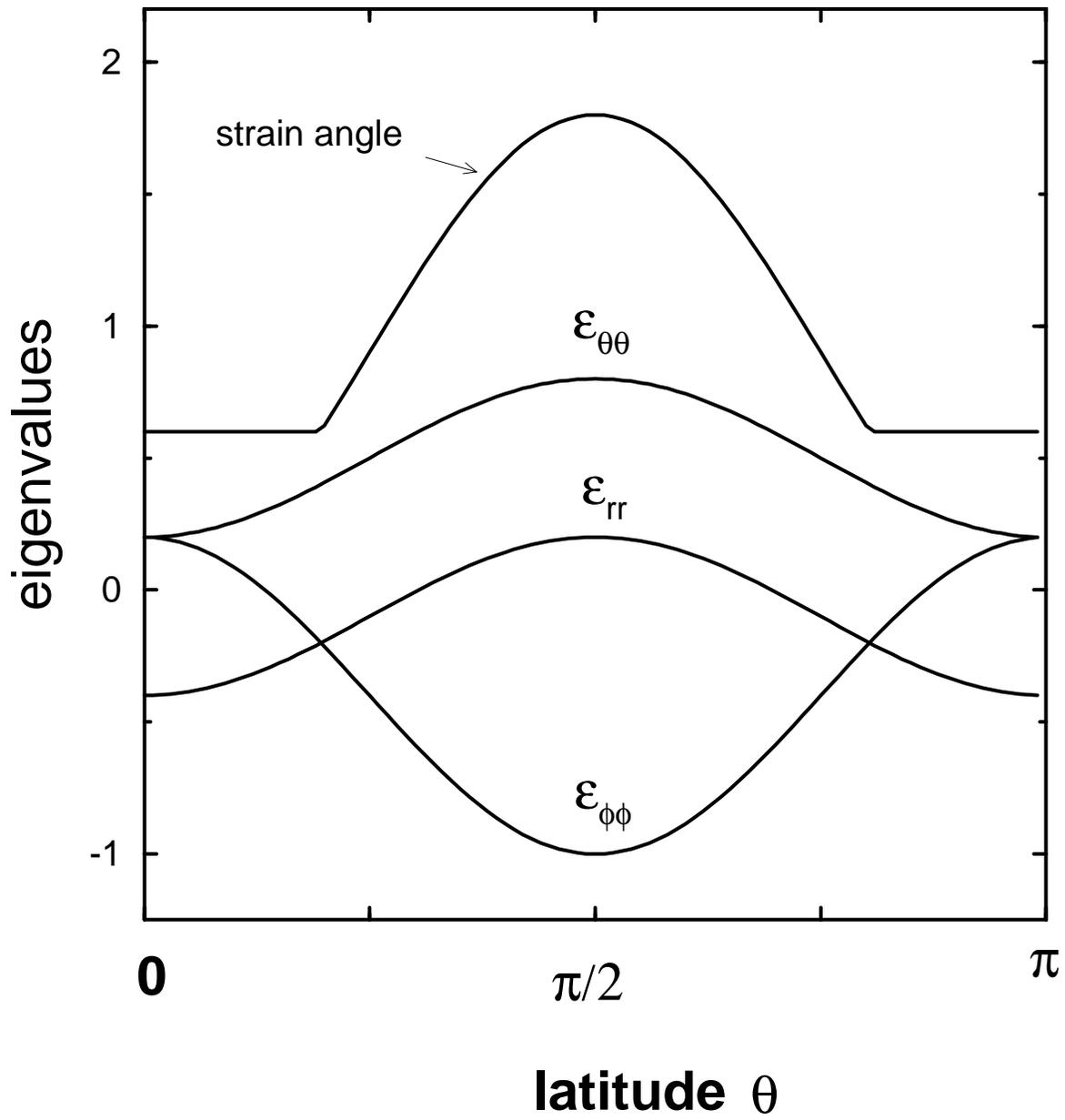
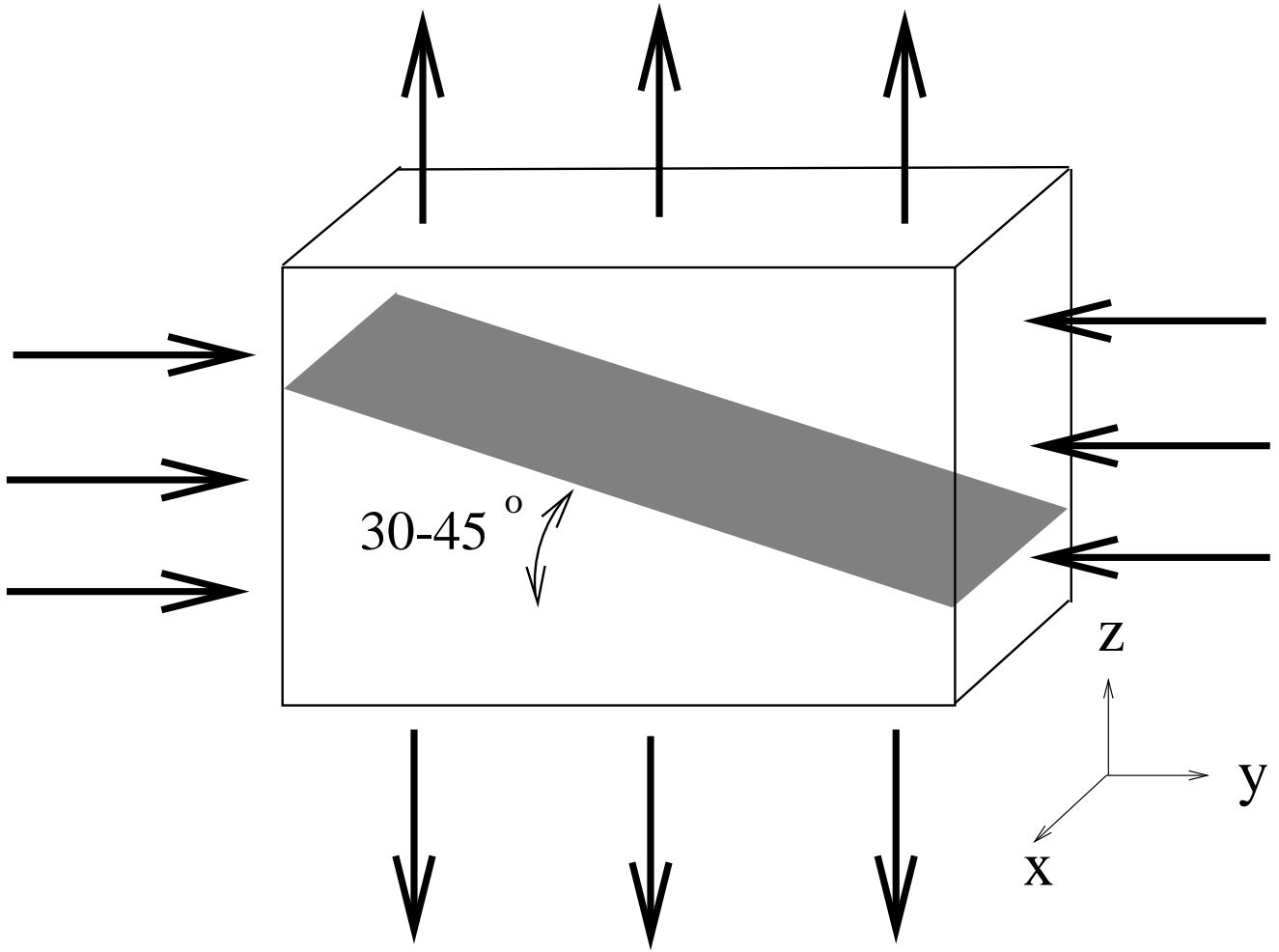
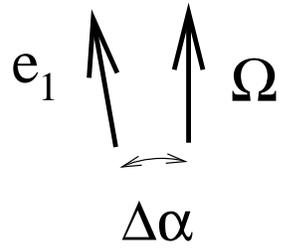
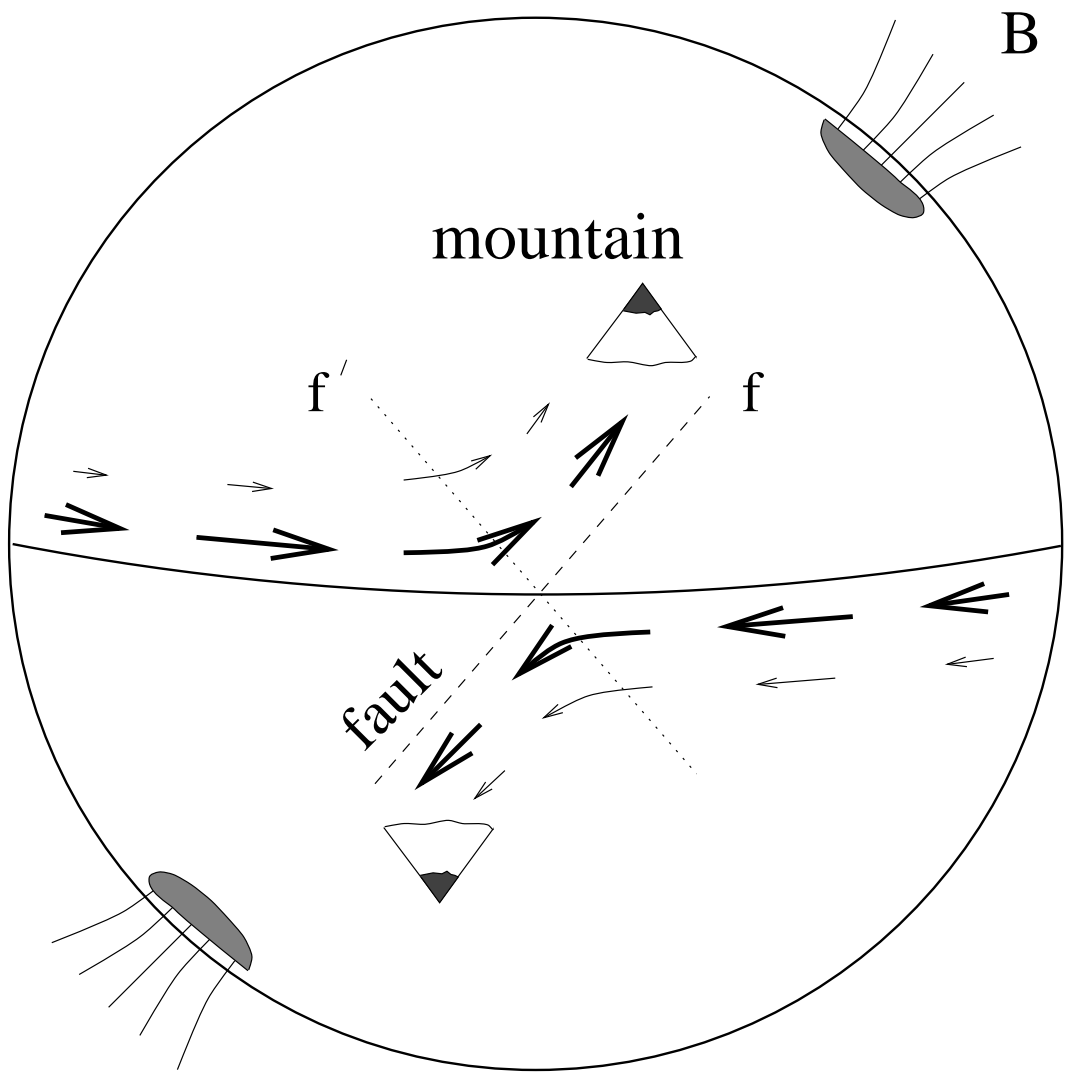
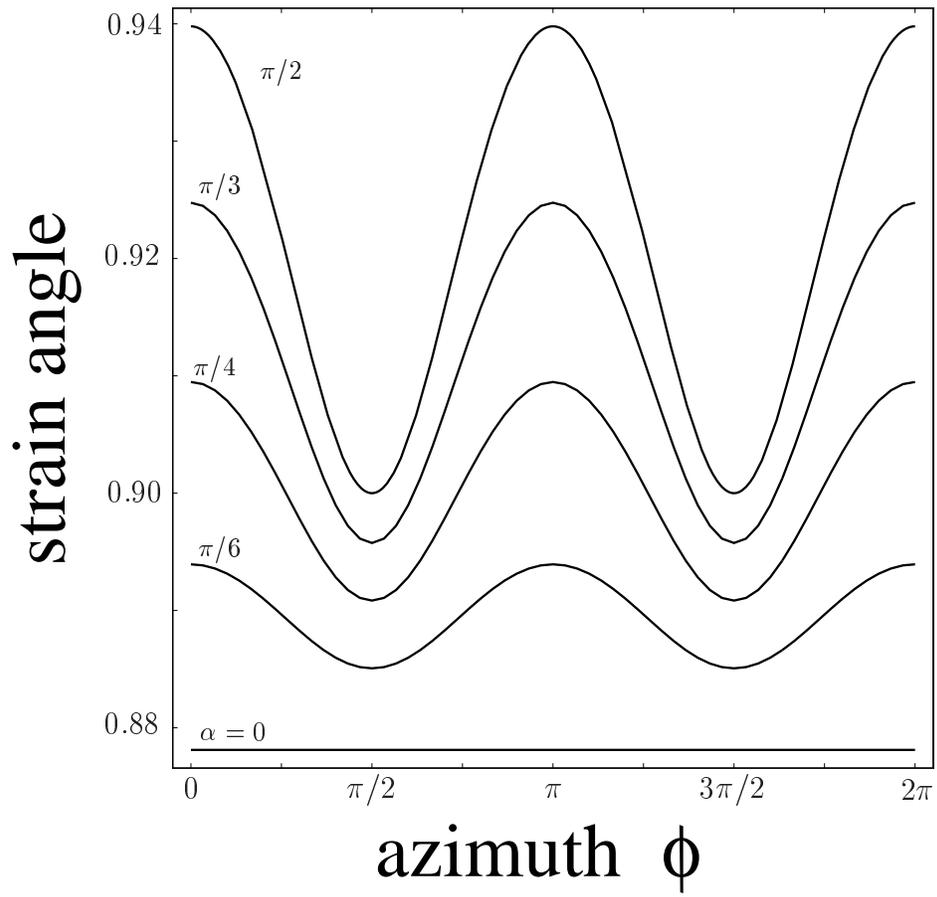
\end{document}